\documentstyle[preprint,prl,aps,psfig]{revtex} 

\newcommand {\be} {\begin{equation}} 
\newcommand {\ee} {\end{equation}}

\begin{document}
\draft

\title{General properties of propagation in chaotic systems}

\author{G. Giacomelli$^{\star\dag}$, 
R. Hegger$^{\ddag,\star}$, A. Politi$^{\star,\dag}$, and
 M. Vassalli$^{\star}$}
\bigskip
\address{
$^\star${\it Istituto Nazionale di Ottica Applicata, L.go E. Fermi 6,
I-50125 Firenze, Italy}\\
$^\dag${\it Istituto Nazionale di Fisica della Materia, Unit\`a di
Firenze}\\
$^\ddag${\it Max-Planck Institut f\"ur Physik Komplexer Systeme, 
Dresden, Germany}
}

\date{\today}
 
\maketitle
\vskip 1 true cm
\begin{abstract}

We conjecture that in one-dimensional spatially extended systems the 
propagation velocity of correlations coincides with a zero of the convective
Lyapunov spectrum. This conjecture is successfully tested in 
three different contexts: (i) a Hamiltonian system (a Fermi-Pasta-Ulam chain 
of oscillators); (ii) a general model for spatio-temporal chaos (the complex 
Ginzburg-Landau equation); (iii) experimental data taken from a $CO_2$ laser 
with delayed feedback. In the last case, the convective Lyapunov exponent 
is determined directly from the experimental data.
\end{abstract}

\vskip 1.cm

Since the recognition that deterministic chaos is an ubiquitous feature of 
nonlinear systems, much has been understood about their dynamical properties.
An important example is the discovery of the relationships between 
Lyapunov exponents, measuring the divergence rate of nearby trajectories, and 
either geometrical or information-theoretic properties such as fractal 
dimensions \cite{KY} and dynamical entropies 
\cite{Pesin}. Nevertheless, little progress has been made to establish a
link between chaotic indicators and directly observable properties. In this 
Letter we show the existence of a general and remarkable connection: the 
velocity of correlation propagation is determined by the vanishing
of the convective Lyapunov exponent.

\begin{figure}
\psfig{file=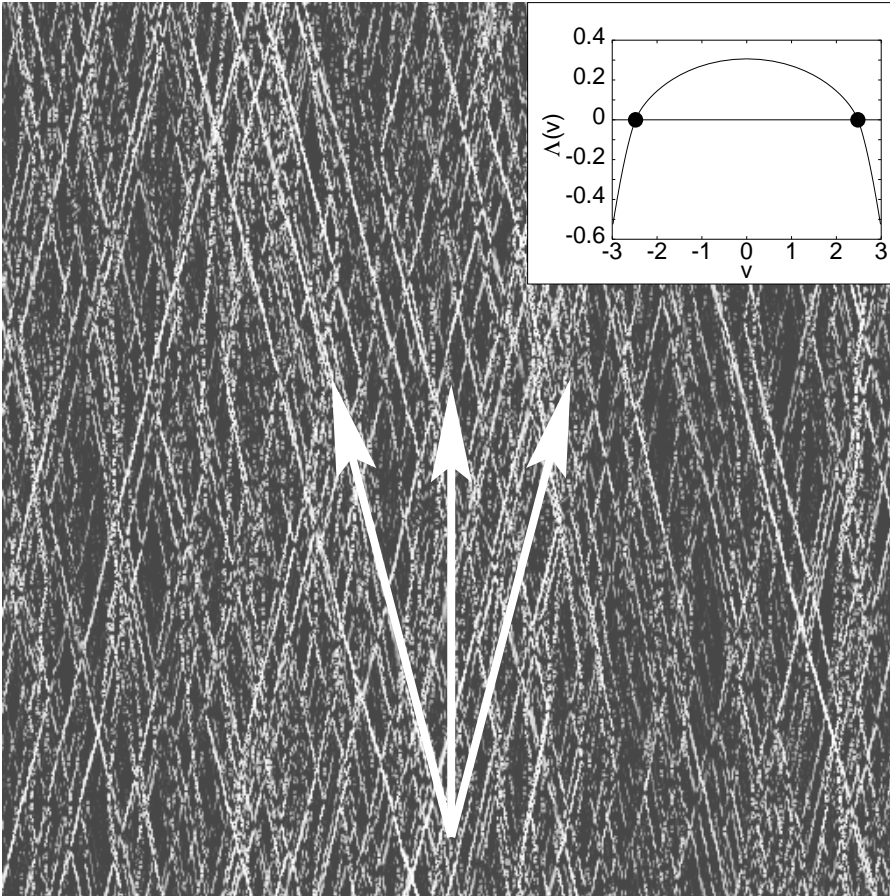,width=12. truecm,angle=270}
\caption{Space-time representation of the heat-flux dynamics in a chain of 
512 FPU oscillators. Time flows from bottom to top as in the other figures.
The total time span is 51.2 in the rescaled units of Eq. (1). The two
tilted arrows correspond to the optimal propagation velocity as determined
from the computation of the space-time correlation function of the heat-flux
(see Eq.~\ref{eq.flux}). The vertical arrow corresponds to the direction
of the maximal growth rate for infinitesimal perturbations. In the inset, the
dependence of the convective Lyapunov exponent is reported as a function 
of the velocity, i.e. the slope in the space-time representation.}
\label{fig.fpu} 
\end{figure}

On the one hand, we know that chaotic systems are generally characterized by 
rapidly decaying correlations. This does not prevent the onset of sizable 
propagation phenomena, as apparent in Fig.~\ref{fig.fpu} where, as an 
example, we report the space-time representation of the local heat-flux in a 
Fermi-Pasta-Ulam (FPU) chain (see 
later for a definition of both the observable and the model). A simple tool 
to pinpoint the existence of travelling processes is the autocorrelation 
function $C(x,t) = \langle w(t'+t,x'+x)w(t',x') \rangle$ (or, equivalently,
the structure function) of some observable
$w(x,t)$, where $x$ and $t$ denote space and time variables, respectively,
and $\langle \cdot \rangle$ denotes a time average (we shall always assume
that ergodicity holds). The most effective propagation processes can then be 
identified by estimating the velocity 
$v =\overline v$ that minimizes the decay of $C(vt,t)$. In many chaotic 
systems the minimum is attained for a non-zero velocity. 

On the other hand, Lyapunov exponents represent the right tool to investigate
the evolution of an infinitesimal perturbation $\delta(x,t)$. If the initial
perturbation $\delta(x,0)$ is localized around the origin, it has been shown
that, in the limit $t \to \infty$,
\cite{DK}
\be
   \delta(x,t)  \simeq  \exp[ \Lambda(v=x/t)t ] ,
\ee
where $\Lambda(v)$ is the so-called convective or 
velocity-dependent Lyapunov exponent. In fact, it expresses the growth 
rate of a localized perturbation when observed in a frame moving with 
velocity $v$. The maximum value of $\Lambda(v)$ coincides with the usual 
maximum Lyapunov exponent: in convectively unstable systems, this occurs 
for a suitable non-zero velocity, while it is located in the orgin 
($v=0$) in spatially symmetric chaotic systems. An example of the typical 
behaviour of $\Lambda(v)$ in this latter context can be seen in the inset 
of Fig.~\ref{fig.fpu}. There, one can 
notice a further general feature: moving away from the maximum, $\Lambda(v)$ 
decreases, becoming negative above some critical velocity $v^*$. 
Therefore, any perturbation ``travelling'' faster than the critical velocity 
is exponentially damped and thus becomes quickly negligible. This fast
damping prevents any coherent propagation of information and, therefore,
we expect that a meaningful long-term coherence cannot be maintained for 
such large velocities. 

In the opposite limit of small velocities, it is well known that an 
exponential separation of trajectories leads to a fast decoherence, since 
nearby orbits rapidly enter different regions of the phase-space. 
Accordingly, strong correlations cannot again be maintained. The only moving 
frame in which neither of the two effects actively contributes to destroying
correlations is precisely that one corresponding to the neutral point $v^*$, 
so that the most effective propagation phenomena should occur exactly at the
velocity $v=v^*$.

The first system where we have tested this conjecture is a chain of FPU
oscillators \cite{FPU}, an idealized microscopic Hamiltonian model 
for an insulating solid, 
\be
\ddot q_i =  F(q_{i-1}-q_i) - F(q_{i}-q_{i+1}) 
\label{eq.fpu}
\ee
where $q_i$ represents the displacement of the $i$th particle from its 
equilibrium position and $F(x) = -x -x^3$ is the force field. This is a 
simple nonlinear model, introduced to test the ergodic hypothesis in the 
first numerical experiment ever performed \cite{FPU}. Within the 
large number of features that have been found while investigating the 
dynamics of the above model, here we are interested in the
propagation phenomena recently discovered in connection with the study of
heat conductivity. In order to clarify the observed anomalous transport
properties, the authors of Ref.~\cite{LLP} have performed microcanonical
simulations (with periodic boundary conditions), monitoring the local heat 
flux
\be
j(t,i)=\frac{\dot q_i}{2} [F(q_{i-1}-q_i)+ F(q_i-q_{i+1})] .
\label{eq.flux}
\ee
From the behaviour of the autocorrelation function
$\langle j(t'+t,i'+i)j(t',i') \rangle$, they clearly found a propagation 
velocity $\overline v \simeq 2.47$, when the energy per particle is $e=8.8$. 
The space-time representation of $j(t,i)$ in Fig.~\ref{fig.fpu} demonstrates 
directly the existence of symmetric propagation phenomena along the 
directions identified by the cross-correlation analysis and denoted by the
tilted arrows in the figure.

For what concerns the spectrum of convective Lyapunov exponents, rather than 
letting an initially localized perturbation evolve, we have preferred to 
follow the procedure devised in Ref.~\cite{PT}, as it suffers much less 
problems of finite-size corrections \cite{LPT1}. The readers interested in a
thorough explanation of the procedure can consult Refs.~\cite{PT,LPT1}; here,
for the sake of completeness, we summarize the key steps.
Very briefly, the method consists in computing the maximum Lyapunov 
exponent $\lambda(\mu)$ of a special class of perturbations: those exhibiting 
an exponential profile with an imposed growth rate $\mu$. This can be done
by linearizing Eqs.~(\ref{eq.fpu}) to obtain the standard equations for a 
generic perturbation $\delta q_i$. By then introducing the Ansatz 
$\delta q_i  = {\overline {\delta q_i}} \exp (\mu i)$, one obtains the
differential equation for the ``envelope'' ${\overline {\delta q_i}}$. The
corresponding growth rate is the generalized Lyapunov exponent 
$\lambda(\mu)$.  The velocity-dependent Lyapunov spectrum $\Lambda(v)$ is 
finally obtained by Legendre transforming $\lambda(\mu)$
($v = d\lambda/d\mu$; $\Lambda = \lambda(\mu) -\mu v$).
{\it Mutatis mutandis}, $\lambda(\mu)/\mu$ can be intepreted as the 
``phase velocity'' of waves with wavelength ``$\mu$'', while 
$v$ can be read as the corresponding group velocity.

The result for the FPU system is reported in the inset of Fig.~\ref{fig.fpu}:
it reveals two symmetric zeros, whose value corresponds to the direction
of the straight lines visible in the corresponding pattern. However,
more importantly, the absolute value of the marginal velocity ($\approx 2.46$) 
is in full agreement with the velocity previously estimated from the 
behaviour of the correlation function. As a first check of the generality of 
this identity, we have repeated the same analysis for different energy 
densities, namely $e=1.$ and $e=0.1$. The data are altogether reported in 
Table 1, where one can see that $\overline v$ always agrees with $v^*$ 
within the numerical error.

\begin{figure}
\psfig{file=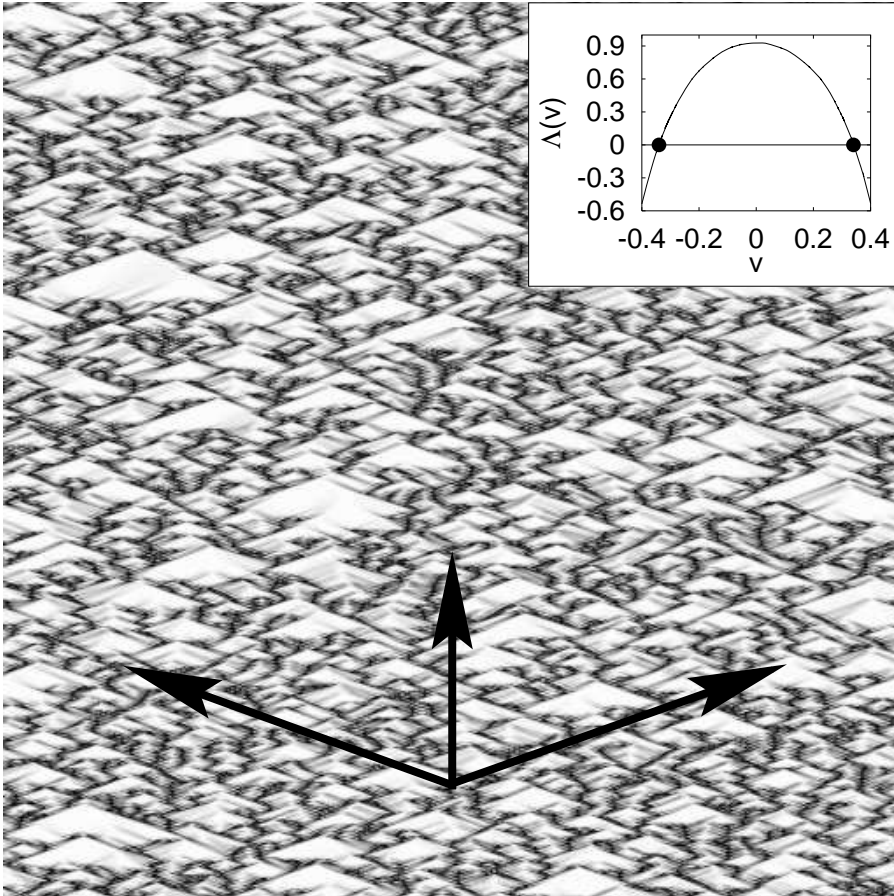,width=12. truecm,angle=270}
\caption{Space-time representation of the square amplitude of the complex
Ginzburg-Landau equation in a system of length 51.2 over a time span equal
to 50. The meaning of the arrows and of the inset is the same as in the 
previous figure.}
\label{fig.cgl} 
\end{figure}

The second model we have considered is the complex Ginzburg-Landau (CGL) 
equation, a partial differential equation describing chaotic properties of 
generic systems close to oscillatory instabilities (it has been derived, 
e.g., in hydrodynamics\cite{gl1} and laser physics\cite{gl2}), 
\be
 \frac{\partial u}{\partial t} = u - (1 - ib) |u|^2u + (1 + ic) 
\frac{\partial^2 u}{\partial x^2}
\label{eq.cgl}
\ee
where $u$ is a complex field which, in general, represents the slowly-varying
amplitude of some relevant mode. For $b=2.3$ and $c = 0$, propagation
phenomena are clearly visible, as it can be noticed in the pattern reported in
Fig.~\ref{fig.cgl} (the equations have been integrated by using the algorithm
described in Ref.~\cite{FGT} and adopting again periodic boundary conditions). 
It is instructive to notice that, at variance with the previous Hamiltonian 
system, now the pattern is no longer invariant under time reversal. This is 
an obvious consequence of the dissipative nature of the CGL equation.
From the computation of the correlation function 
$\langle |u|^2(t+t',x+x')|u|^2(t',x')\rangle$, we can determine the optimal 
propagation velocities which are again symmetric (see the two arrows in 
Fig.~\ref{fig.cgl}) and practically coincide with the velocity of the dark 
structures visible in the pattern. 

The comoving exponents can be again computed by linearizing Eq.~(\ref{eq.cgl})
and assuming an exponential profile for the perturbation. The resulting 
spectrum is reported in the inset of Fig.~\ref{fig.cgl}. It is
symmetric (because of the left-right symmetry in the model) and the zeros of 
the spectrum coincide with the propagation velocities of correlations
(see Table 1) represented by the two tilted arrows in the same figure. 

\begin{figure}
\psfig{file=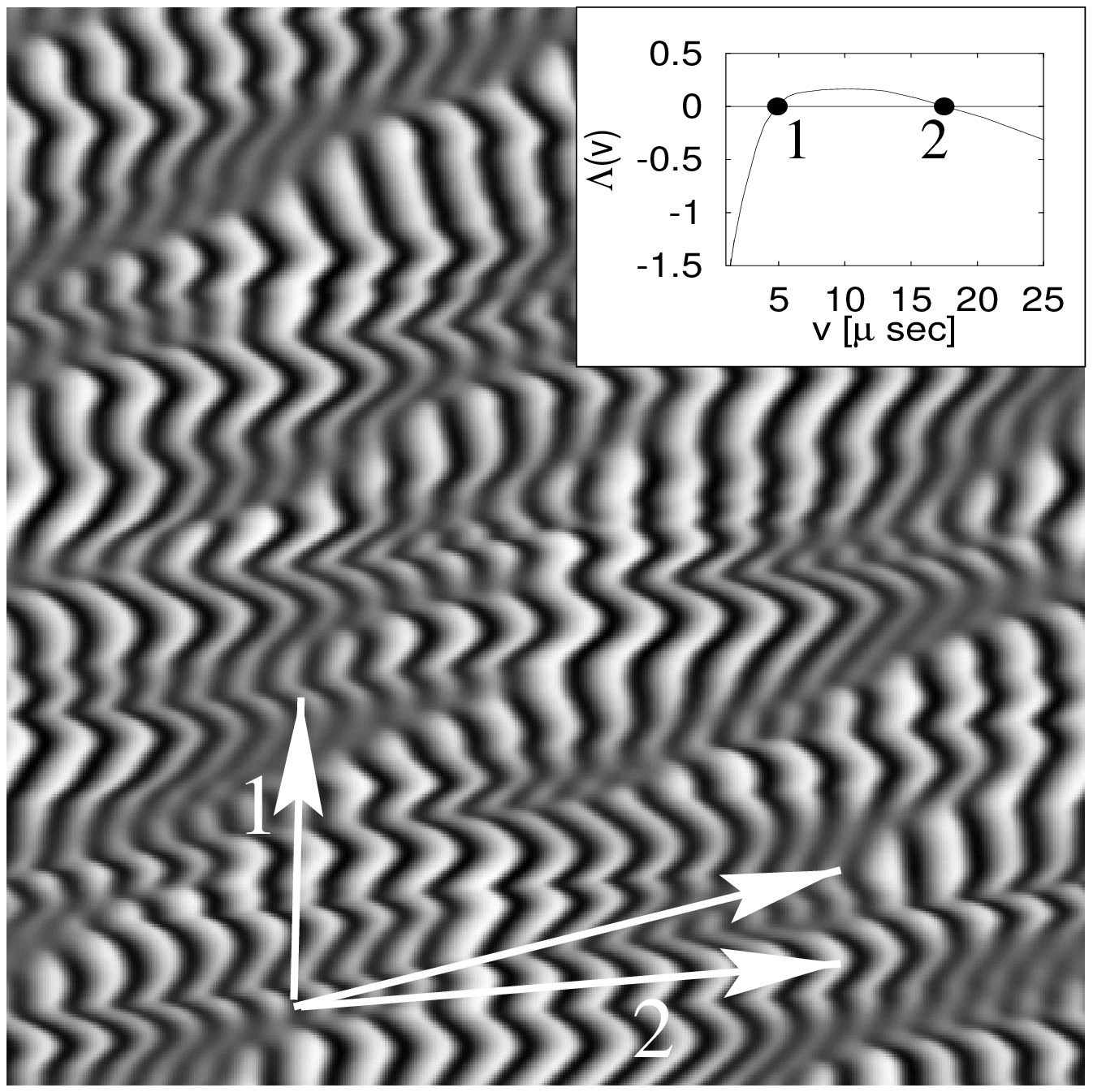,width=12. truecm,angle=270}
\caption{Space-time representation of the field intensity in the $CO_2$
laser experiment described in the text. The horizontal axis represents
the time variable within a delay unit of length 406 $\mu sec$, while the $y$ 
direction corresponds to different units (see the text for a more detailed 
explanation). The central arrow again denotes the direction characterized
by the maximal growth rate.}
\label{fig.las} 
\end{figure}

The last system where we have tested our conjecture, is an experimental one,
namely, a $CO_2$ laser with delayed feedback. This is a physical system that 
has been used in the past to investigate several instabilities both in the 
regime of low and high-dimensional chaos \cite{las1,las2}. Although this is
not, strictly speaking, a spatially extended system, it can be interpreted as
such. The idea consists in decomposing the time variable as $t = n + s \tau$
\cite{AGLM}, where the parameter $\tau$ is close to the delay time (fixed 
equal to $400 \mu s$ in the experiment), 
$n = {\rm Int}(t/\tau)$ is the new, discrete, time variable, and $s$
($0<s<\tau$) is a space-like variable. In a sense, this system is complementary
to a chain of oscillators, since the discrete and continuous character of
time and space axes are exchanged. The validity of this decomposition has been
established in Ref.~\cite{GP}; here, the reader can appreciate its 
meaningfulness by looking at the space-time representation of the intensity of 
the emitted radiation in Fig.~\ref{fig.las}. While looking at the pattern, it 
is important to realize that the slope of the various coherent structures 
depends on the value of $\tau$ adopted in the time decomposition. Here, for 
the sake of clarity, we have adjusted $\tau$ in such a way that the 
pseudo-rolls, responsible for the optimal correlations, are almost vertical 
(see arrow 1). A further important difference with the previous models is 
the absence of left-right symmetry. Even more, causality implies that no 
leftwards (negative) propagation is possible at all. 

Since no theoretical model exists which reproduces the laser dynamics with a
sufficient accuracy, one must exclusively rely on the experimental data. 
However, it has been recently developed an approach to reconstruct the 
dynamics of delayed systems, even when it is so high-dimensional that the 
standard embedding technique is bounded to fail \cite{var,new1}. In short,
given a time-series $y_n$, the method consists in constructing an 
embedding space composed of two-window vectors, 
\be
v_n \equiv (y_n,y_{n-1},\ldots,y_{n-m+1},y_{n-T},\ldots,y_{n-T-m+1}) ,
\ee
where the distance $T$ between the two windows coincides with the delay, while
the length $m$ is the effective number of variables involved in the dynamics. 
Once the proper values of both $T$ and $m$ have been identified
(by minimizing the forecast error), a local linear model can be constructed, 
by fitting the behaviour in a suitable neighbourhood of each point in the 
embedding space. In the case of the $CO_2$ laser, $m=5$ 
guarantees an excellent reproduction of the original dynamics \cite{new1}.
From the knowledge of the empirical model, one can again compute the convective 
Lyapunov exponents by going through the same intermediate steps. The 
corresponding spectrum is reported in the inset of Fig.~\ref{fig.las}
(notice that, given the peculiarity of the space-time reconstruction, 
velocities are here expressed in time units - more precisely $\mu s$ per 
number of delay units), where we see that it is 
restricted to the positive-$v$ domain (see Ref.~\cite{GP}). As a consequence 
of this asymmetry, both zeros are positive. Thus, in principle, one might 
expect two different propagation velocities. The correlation analysis of the 
experimental data has instead revealed a single velocity, which coincides 
with the smallest zero of the convective Lyapunov spectrum (see Table 1). 
We cannot rule out the possibility that a second, much less effective, 
propagation exists with a larger velocity in coincidence with the second 
zero. However, even if this is not the case, we can at least state that 
whenever propagation of correlations can be inferred from the evolution
of some observable, it must coincide with a netrually stable velocity for
the spectrum of convective Lyapunov exponents. What are the (possibly) 
additional ingredients ensuring the actual existence of propagation 
phenomena and determining their strength seems to be a much harder problem 
that we plan to attack in the future.

We conclude, by restating that we have found a clear evidence of a strict
link between a ``large''-scale feature like the optimal propagation of 
correlations and the evolution of infinitesimal perturbations, a problem 
which, because of its very nature, can be formulated in terms of Lyapunov
exponents. Accordingly, albeit deterministic chaos contributes to a fast 
decay of correlations in low-dimensional systems, it is compatible with 
propagation phenomena as soon as spatial degrees of freedom come into play. 

Financial support from the European Union (contract N. PSS 1043) is
acknowleged.

\begin{table}
\begin{center}
\begin{tabular}{|c|c|c|}
System &  $v^*$  & $\overline v$\\
\hline
FPUa & $2.46 \pm 0.01$ & $2.47 \pm 0.04$\\
FPUb & $1.55 \pm 0.01$ & $1.53 \pm 0.05$\\
FPUc & $1.11 \pm 0.01$ & $1.2 \pm 0.15$\\
CGL  & $0.38 \pm 0.01$ & $0.37 \pm 0.02$\\
Laser & $(4.9 \pm 0.1)\mu$ sec & $(4.8 \pm 0.2)\mu$ sec \\
\hline
\end{tabular}
\end{center}
\caption{The velocity $v^*$ of the zero convective Lyapunov exponent
and the velocity $\overline v$ of the optimal propagation of correlations
in an FPU chain (for $e=8.8$ (a), $e=1$ (b), and $e=0.1$ (c)), in the
CGL equation, and in the $CO_2$ laser system (from experimental data). 
Except for the last case, velocities are expressed in adimensional units.} 

\label{table1}
\end{table}

\begin{thebibliography}{99}

\bibitem{KY} J.L. Kaplan  and J.A. Yorke, Lect. Notes in Math. {\bf 730},
228 (1979).
\bibitem{Pesin} Ya.B. Pesin Ya.B., Russ. Math. Surv. {\bf 32} 55 (1977).
\bibitem{DK} R.J. Deissler and K. Kaneko, Phys. Lett. {\bf 119} 397 (1987).
\bibitem{FPU} E. Fermi, J. Pasta, and S. Ulam, 
Los Alamos Science Laboratory Report No. LA-1940 (1955) later
published in E. Segr\'e (Ed.), Collected Papers of Enrico Fermi, vol. 2, 
University of Chicago Press, Chicago, 1965, p.978.
\bibitem{LLP} S. Lepri, R. Livi, and A. Politi, Europhys. Lett. {\bf 43}, 
271 (1998).
\bibitem{PT} A. Politi and A. Torcini, CHAOS, {\bf 2} 293 (1992).
\bibitem{LPT1} S. Lepri, A. Politi, and A. Torcini, J. Stat. Phys. {\bf 82}, 
1429 (1996).
\bibitem{gl1} P. Manneville, {\it Dissipative structures and weak turbulence},
(Academic Press, New-York 1990).
\bibitem{gl2} A.C. Newell and J.V. Moloney, {\it Nonlinear Optics}, 
(Addison-Wesley Pub., Redwood City 1992).
\bibitem{FGT}A. Torcini, H. Frauenkron, and P. Grassberger, Phys. Rev. E, 
55 5073 (1997). 
\bibitem{las1} F.T. Arecchi, W. Gadomski, and R. Meucci,
  Phys. Rev. A {\bf 34}, 1617 (1986).    
\bibitem{las2} G. Giacomelli, R. Meucci, A. Politi, and F.T. Arecchi,
Phys. Rev. Lett. {\bf 73}, 1099 (1994).    
\bibitem{AGLM} F.T. Arecchi, G. Giacomelli, A. Lapucci, and R. Meucci,
Phys. Rev. {\bf A 45}, 4225 (1992).
\bibitem{GP} G. Giacomelli and A. Politi, Phys. Rev. Lett. {\bf 76}, 
2686  (1996).    
\bibitem{var} R. Hegger, M.J. Buenner, H. Kantz, and A. Giaquinta,
Phys. Rev. Lett. {\bf 81}, 558 (1998).
\bibitem{new1} M.J. B\"unner, M. Ciofini, A. Giaquinta, R. Hegger, H. Kantz, 
R. Meucci, and A. Politi, unpublished.

\noindent

\end{thebibliography}
\end{document}